\documentclass[preprint,showpacs,preprintnumbers,amsmath,amssymb]{revtex4}
\usepackage{graphicx,color}

\begin{document}

\title{Temporal response to harmonic driving in electroconvection}

\author{Tibor T\'oth-Katona, N\'{a}ndor \'{E}ber and
\'{A}gnes Buka} \affiliation{Research Institute for Solid State
Physics and Optics, Hungarian Academy of Sciences, H-1525 Budapest,
P.O.B. 49, Hungary}
\date{\today}

\begin{abstract}
The temporal evolution of the spatially periodic electroconvection
(EC) patterns has been studied within the period of the driving {\it
ac} voltage by monitoring the light intensity diffracted from the
pattern. Measurements have been carried out on a variety of nematic
systems, including those with negative dielectric and positive
conductivity anisotropy, exhibiting "standard EC" (s-EC), those with
both anisotropies negative exhibiting "non-standard EC" (ns-EC), as
well as those with the two anisotropies positive. Theoretical
predictions have been confirmed for stationary s-EC and ns-EC
patterns. Transitions with Hopf bifurcation have also been studied.
While traveling had no effect on the temporal evolution of
dielectric s-EC, traveling conductive s-EC and ns-EC patterns
exhibited a substantially altered temporal behavior with a
dependence on the Hopf frequency. It has also been shown that in
nematics with both anisotropies positive, the pattern develops and
decays within an interval much shorter than the period, even at
relatively large driving frequencies.

\end{abstract}
 \maketitle

\section{\label{sec:intro}Introduction}

Electroconvection (EC) in a thin layer of a nematic liquid crystal
(NLC) is a paradigm of pattern forming instabilities \cite{kramer,
Pesch}, and as such, it has been widely studied both
experimentally and theoretically.

These studies have concluded that the sign (and magnitude) of the
anisotropies of the dielectric permittivity,
$\epsilon_a=\epsilon_{||}-\epsilon_{\perp}$, and of the electrical
conductivity, $\sigma_a=\sigma_{||}-\sigma_{\perp}$, and the initial
director alignment are the key parameters that determine whether,
and what kind of EC patterns can exist \cite{Kluw}. Subscripts
$_{||}$ and $_{\perp}$ indicate here the values measured parallel
with and perpendicular to the initial director orientation {\bf n},
respectively. Due to the crucial role of the anisotropies it is
convenient to classify NLCs into $(-,+)$, $(-,-)$, $(+,+)$, and
$(+,-)$ compounds, with the first sign corresponding to $\epsilon_a$
and the second one to $\sigma_a$.

Electroconvection typically appears at a threshold rms value $U_c$
of the driving ac voltage of frequency $f$ as a spatially periodic
pattern (a set of rolls) characterized by a wave vector
$\textbf{q}_c$. In most cases the pattern involves a distortion of
the director field accompanied with charge separation and vortex
flow, forming a feedback loop known as the Carr-Helfrich mechanism
\cite{kramer,Kluw}. Its theoretical formulation, the so called
standard model (SM) of EC \cite{Bod,kramer} provides a
quantitative description of basic pattern characteristics -- such
as $U_c(f)$ and $\textbf{q}_c(f)$ -- for various combinations of
the key parameters where EC can occur either as a direct
transition (planar $(-,+)$\cite{kramer} or homeotropic $(+,-)$
\cite{buka}) or as a secondary instability (e.g., homeotropic
$(-,+)$ \cite{Kai96,Buka00}). Therefore these patterns will be
referred to as standard EC (s-EC) patterns further on. It has to
be mentioned, however, that according to the rigorous analysis by
SM, the boundaries of existence of s-EC modes in the
($\epsilon_a$, $\sigma_a$) space do not coincide exactly with the
sign inversion of the anisotropies \cite{Kluw}.

Though the SM predicts a bifurcation to stationary EC patterns,
occasionally (for some range of material and control parameters
$\sigma _{\perp}$, $f$ and sample thickness $d$) traveling rolls
(TRs) have also been observed experimentally at threshold in various
NLCs \cite{Kai78,Joets88, Rehberg89,Dennin95} indicating a
Hopf-bifurcation. In order to explain these features, a theoretical
extension of the SM (the weak electrolyte model -- WEM) has been
developed by taking into account diffusion, recombination and
dissociation of ionic charge carriers \cite{Treiber95}. The $U_c(f)$
and $\bold{q_c}(f)$ dependencies calculated from the WEM do,
however, practically coincide with those provided by the SM.

According to SM, the existence of s-EC patterns is excluded for
$(-,-)$ nematics \cite{Kluw,de Gennes}. However, convection in ac
electric field associated with roll formation has long ago been
observed, e.g., in homologous series of
N-(p-n-alkoxybenzylidene)-n-alkylanilines, or in
di-n-4-4'-alkyloxyazoxybenzenes \cite{leger, blinov}, and recently
in 4-n-alkyloxy-phenyl-4-n'-alkyloxy-benzoates \cite{Ela,Toth07}.
The characteristics of these patterns, like orientation of the
rolls, contrast, frequency dependence of $q_c$ and $U_c$, and the
director distribution in space and time, are different from those of
the s-EC. Since this roll formation process falls outside of the
frame of the SM it has been called non-standard electroconvection
(ns-EC) \cite{Ela}.

After several attempts of explanation
\cite{leger,blinov,blinov77,madu}, the scenario has recently been
understood, and experimental results have been quantitatively
reproduced by the extended SM which incorporates flexoelectricity
\cite{Krekhov08}. Though in this scenario the standard Ohmic charge
separation of the Carr-Helfrich mechanism leads to a negative
feedback due to $\sigma_a < 0$, this effect can be compensated and
the homogeneous state is destabilized by the flexoelectric charges.

There is a number of experimental evidences
\cite{blinov77,Blinov,Trufanov,Rout,Rjumtsev93,Nakagawa,Kumar2010}
that EC may occur in $(+,+)$ nematics too, which also infringes the
predictions of the (extended) SM. The isotropic (Felici-Benard)
mechanism \cite{felici,Blinov} has been proposed to explain the
observed instabilities \cite{blinov77,Trufanov}, however, a rigorous
theoretical formulation of this mechanism, which is capable to
provide the main pattern characteristics, is unfortunately not yet
available.

In the present paper we intend to focus on the time evolution of the
EC patterns, hence one has to take into account the characteristic
time scales involved in EC. According to SM the Carr-Helfrich
mechanism has three relevant time scales: the director relaxation
time $\tau _{\rm d} = \frac{\gamma _1 d^2}{K_{1} \pi ^2}$, the
charge relaxation time $\tau _{\rm q} = \frac{ \epsilon _0 \epsilon
_{\perp}}{\sigma _{\perp}}$ and the viscous relaxation time
$\tau_{\rm v}= \frac{\rho d^2}{\alpha_4/2}$. Here $K_{1}$ is the
splay elastic modulus, $\rho$ is the density, $\gamma _1$ is the
rotational, and $\alpha_4/2$ is the isotropic viscosity, $d$ is the
sample thickness. Typically, $\tau _{\rm d} \gg \tau _{\rm q} \gg
\tau_{\rm v}$ with $\tau_d =\mathcal{O}(1$s) \cite{kramer}. $\tau
_{\rm q}$ affects the so-called cut-off frequency $f_c$ which
separates two basic regimes of s-EC patterns having opposite time
symmetries. Below $f_c$ -- in the conductive regime -- the charge
distribution oscillates with the applied ac voltage, while above
$f_c$ -- in the dielectric regime -- the director and the flow
follows the electric field. $\tau _{\rm d}$, together with $q_{\rm
c}$ and the voltage deviation from $U_{\rm c}$, determines how fast
the pattern develops (decays) after switching on (off) the driving
voltage. It has been shown recently that the decay rates grow
strongly with increasing $q_{\rm c}$ \cite{Eber04,Pesch06}. In case
of traveling waves further time scales enter the game related to the
ionic processes as described by the WEM; that makes the appearance
of Hopf bifurcation and the calculation of the Hopf frequency $f_H$
possible \cite{Treiber95}.

The period $t_0=1/f$ of the driving voltage corresponds to an
additional characteristic time. In most cases $\tau_{\rm d} >> t_0
>> \tau_{\rm v}$ holds -- as it is in our experiments reported later in
the paper too. We note here that if $t_0$ becomes comparable or
larger than $\tau_{\rm d}$ (low frequency driving), the time
evolution of the pattern may suffer a significant change; the
pattern might exist only as bursts in a part of the period $t_0$ as
proven recently for dielectric s-EC \cite{May08}.

In regard of the experimental methods, most of the EC measurements
involve primarily optical microscopy. Since the driving frequencies
in EC studies are mostly comparable with, or larger than the
standard video rates, microscopy usually cannot provide information
on the temporal evolution of the director within one period of the
applied AC voltage unless stroboscopic techniques
\cite{Schneider92,Messlinger07}, or high-speed cameras are used.

Light diffraction may serve as an alternative technique. A
comprehensive theoretical analysis of diffraction gratings formed by
EC has been developed \cite{Kosmo87,Zengino88,Zengino89,Pesch06}.
Diffraction experiments were mainly devoted to the laser diffraction
efficiency of EC \cite{Caroll72}, studying the cascade of EC
patterns above $U_c$ \cite{Lu71,Xu07}, measuring the frequency of
the oscillatory motion of Williams domains above $U_c$
\cite{Akahoshi76,Miike84}, investigating stochastically excited EC
\cite{John02,John03,John04}, and determining the growth and decay
rates of EC patterns \cite{Papa99,John03,Eber04,Pesch06}.

Light diffraction is especially an adequate technique for studying
the temporal evolution as the intensity of the first order
diffraction fringes $I_{\pm 1}$ is directly related -- near the
onset -- to the square of the director distortion angle $\theta$
($I_{\pm 1} \propto \theta^2$). Interestingly, however, we are not
aware of any systematic investigation of the light intensity
diffracted from EC patterns at $U_c$ within the period of the
driving voltage, except of the early work \cite{Orsay71} where a
single screenshot of synchronous oscilloscopic recording of the
excitation voltage and of the light intensity scattered from
dielectric s-EC pattern is presented.

In this paper we intend to fill this gap. After introducing the
experimental set-up and the NLCs used for the measurements in
Sec.~\ref{sec:setup}, in Sec.~\ref{sec:exp} we present our
experimental results on the temporal variations of the light
intensity diffracted from various EC patterns at (or close to) the
onset of the instability. On the one hand, our motivation was to
confirm/test the theoretical predictions of the SM. Hence in
Sec.~\ref{subsec:stacsEC} we present the temporal evolution of the
diffracted light intensity for stationary conductive and dielectric
s-EC in $(-,+)$ compounds. On the other hand, for several other
pattern morphologies unfortunately no such predictions exist, so
experimental determination of the behaviour is of high importance.
For example, in the case of traveling s-EC patterns, which are
discussed in Sec.~\ref{subsec:travsEC}, the WEM should presumably
provide the temporal evolution of the director distribution.
Measurements have also been carried out in NLCs exhibiting ns-EC. In
Sec.~\ref{subsec:8_7nsEC} time dependence of ns-EC patterns of a
$(-,-)$ nematic is addressed. For this case the extended SM foresees
a time symmetry similar to that of the dielectric s-EC
\cite{Krekhov08}. Measurements presented in Sec.~\ref{subsec:5CB}
for EC in $(+,+)$ NLCs may serve as important ingredients for a
future construction of a theoretical description. Finally, in
Sec.~\ref{sec:discussion} we summarize the conclusions obtainable
from our measurements and from the comparison with existing
theories.

\section{\label{sec:setup}Substances and experimental set-up}

Substances for the present studies have been carefully selected to
cover all combinations of $\epsilon_a$ and $\sigma_a$, but the
$(+,-)$ case which is not considered here. For investigating both
conductive and dielectric s-EC in $(-,+)$ materials, we have used
nematic mixtures Phase 4, Phase 5 and Phase 5A (from Merck Co.).
In these systems both stationary and traveling s-EC patterns could
be realized at a fixed temperature of $T=30^{\circ}$C. For
studying ns-EC in $(-,-)$ nematics, two members of the homologous
series of 4-n-alkyloxy-phenyl-4-n'alkyloxy-benzoates
\cite{Kresse80} (labelled as \textbf{n/m}, where $n$ and $m$ are
the alkyl-chain lengths of the tails of the molecules) have been
used. We have reported recently \cite{Ela,Toth07} that, while
their dielectric anisotropy is always negative, the two selected
members of the series -- namely \textbf{8/7} and \textbf{10/6} --
exhibit sign inversion of $\sigma_a$ at some temperature in the
nematic phase, and therefore, one can conveniently study both s-EC
[at temperatures where the compounds are of $(-,+)$ type] and
ns-EC [at temperatures where they are of $(-,-)$ type]. It is
convenient to introduce here a reduced temperature as
$T^{*}=(T-T_{NS})/(T_{NI}-T_{NS})$, with $T_{NI}$ and $T_{NS}$
denoting the nematic--isotropic and the nematic--smectic phase
transition temperatures, respectively. This reduced temperature
will be used throughout the paper for compounds \textbf{8/7} and
\textbf{10/6}. Finally, as a $(+,+)$ nematic, commercially
available 4-cyano-(4'-pentyl)biphenyl (5CB) has been studied at
$T=30^{\circ}$C.

NLCs have been enclosed between two parallel glass plates coated
with etched transparent indium tin oxide (ITO) electrodes and with
rubbed polyimide to assure planar initial alignment. Cells of
different thickness in the range of $3 \mu$m $\leq d \leq 20 \mu$m
have been used. The direction of the director at the surfaces is
chosen as the $x$-axis. A sinusoidal electric voltage of frequency
$f$ and amplitude $\sqrt{2} U$ provided by a function generator was
applied across the sample (along the $z$-axis) via a high voltage
amplifier.

For electroconvection measurements the cells have been placed into
an oven with thermal stability within $\pm 0.05 ^{\circ}$C. EC
patterns have been studied with two different techniques: polarizing
microscopy and laser diffraction. Microscopic observations, either
by the shadowgraph (single polarizer) technique or with two crossed
(or nearly crossed) polarizers, have been used here to verify the
type of the EC pattern only. In the diffraction set-up a central
area of about 1mm$\times 2$mm of the cell has been illuminated with
a beam of a laser-diode of wavelength $\lambda = 650$ nm. Due to the
periodicity of the EC patterns a sequence of diffraction fringes
(far field image) appear on a screen placed normal to the initial
beam at a distance of $L = 0.660 $m from the sample. An optical
fiber (with a diameter of 1 mm) has been positioned onto the center
of a first order diffraction spot. The fiber transmitted the
diffracted light into a photo-multiplier working in its linear
regime. The output of the photo-multiplier has been fed through a
current-to-voltage converter into a digital oscilloscope. That has
allowed computer recording of the fringe intensity $I_{\pm 1}$ at 8
bit resolution with adjustable sampling rate.

We have monitored $I_{\pm 1}(t)$ in order to
check whether there are modulations of the intensity with the
driving frequency or its upper harmonics. In general, at
(or slightly above) the onset the $I_{\pm 1}(t)$ signal mainly
corresponds to a dc level with a superposed second harmonic
component. Therefore, throughout the paper we characterize the intensity
modulation by the ratio $A/B$, where $A$ is the amplitude of the $2f$
component while $B$ is the average intensity with respect to a background dark level
measured at the fringe location for $U\ll U_c$.

\section{\label{sec:exp} Results}

\subsection{\label{subsec:stacsEC}Light diffraction on stationary s-EC patterns}

We start from the case of stationary conductive and dielectric s-EC
where the theoretical expectations given by the SM are firmly
established. The SM provides the temporal variation of the director
field at onset in a (harmonic) Galerkin expansion. One of the basic
differences between the two types of s-EC patterns -- the conductive
and the dielectric rolls -- lies in their time symmetry. According
to the SM, the time dependence of the director is composed of the
odd terms of the expansion ($\theta = C_1 sin(\omega t)+ C_3 sin(3
\omega t) + h.o.t.$) for the dielectric and of the even terms
($\theta = C_0 + C_2 sin(2 \omega t) + h.o.t.$) for the conductive
rolls ($\omega = 2 \pi f$). Focusing on the leading terms only, it
means that in the dielectric regime the director oscillates with the
frequency of the driving ac voltage (and thus passes the $\theta =
0$ state twice in each period), contrary to the case of the
conductive regime, where $\mathbf{n}$ is stationary. Numerical
calculations show that higher harmonics in the Galerkin expansion
have typically a very small weight; even the term next to the
leading one is less than a few percent ($C_2 \ll C_0, C_3 \ll C_1$).
As the intensity of the first order diffracted fringe is $I_{\pm 1}
\propto \theta^2$, one expects $I_{\pm 1} \propto \frac{1}{2} C_1^2
- \frac{1}{2} C_1^2 cos(2 \omega t)+ h.o.t.$ (in terms of
definitions given in Sec.~\ref{sec:setup} $A=B=\frac{1}{2} C_1^2$,
i.e. $A/B=1$) for the dielectric pattern, while for the conductive
one $I_{\pm 1} \propto C_0^2 + 2 C_0 C_2 sin(2 \omega t)+ h.o.t.$
(i.e. $A=2C_0C_2$, $B=C_0^2$, and $A/B \ll 1$) is anticipated.

\begin{figure}[h!t]
\includegraphics[width=8cm]{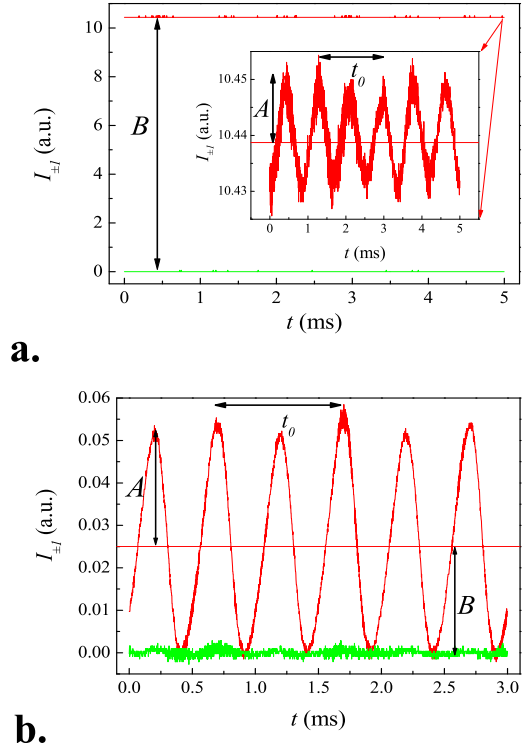}
\caption{(Color online) Temporal evolution of the first order fringe
intensity just above the onset of instability in samples exhibiting
stationary s-EC patterns. The straight line shows the average
intensity, the nearly straight, noisy line at the bottom
(fluctuating around $I_{\pm 1}=0$) corresponds to the background
dark level. The oscillation occurs with twice the driving frequency.
$A$ is the amplitude of the oscillations, $B$ the average intensity
with respect to the dark level, and $t_0$ is the period of the
driving frequency. (a) Phase 5A, $d=12\mu$m, $T=30^{\circ}$C,
$f=600$Hz, stationary conductive s-EC, $A/B \approx 0.001$, the
inset is the blow-up of the intensity oscillations; (b) Phase 5,
$d=12\mu$m, $T=30^{\circ}$C, $f=1$kHz, stationary dielectric s-EC,
$A/B=1$.}\label{fig:stacsEC}
\end{figure}

In order to verify the above predictions diffraction measurements
have been carried out on the nematic liquid crystals Phase 5 and
Phase 5A which are often used for studying s-EC. In Phase 5A only
stationary conductive s-EC rolls have been detected which are
characterized by an almost constant fringe intensity (in general,
$A/B<0.01$) as shown in Fig.~\ref{fig:stacsEC}a. In Phase 5, above
the cut-off frequency $f_c$ stationary dielectric rolls are present
with the fringe intensity falling down to zero twice in each period
of the applied voltage as demonstrated in Fig.~\ref{fig:stacsEC}b
(i.e. $A/B=1$). Therefore, experimental results on the stationary
s-EC patterns of Phase 5 or 5A agree fully with the predictions of
the SM.

\subsection{\label{subsec:travsEC}Light diffraction on traveling s-EC patterns}

While stationary s-EC patterns can fully be described by SM,
traveling of the patterns cannot be explained within its framework.
It could be accounted for by an extension of the SM, considering the
ionic processes of electrical conductivity (the weak electrolyte
model). The WEM inherits the basic Carr-Helfrich mechanism of the EC
pattern formation from the SM and it has been shown that the
frequency dependences of $U_\mathrm{c}$ and $q_\mathrm{c}$ provided
by the two models practically coincide \cite{Treiber95}. Though the
WEM has been analyzed for the Hopf frequency (i.e. traveling), it
has not been done -- to our knowledge -- for the spatio-temporal
distribution of the director. On the one hand, the identity of the
basic instability mechanisms and threshold features might suggest
that $\theta (t)$ for traveling s-EC rolls should be similar to that
of stationary s-EC (which was discussed in
Sec.~\ref{subsec:stacsEC}). On the other hand, the governing
equations of the WEM are more complex than those of the SM, and this
might show up in the temporal behaviour of their solutions.

\begin{figure}[h!t]
\includegraphics[width=8cm]{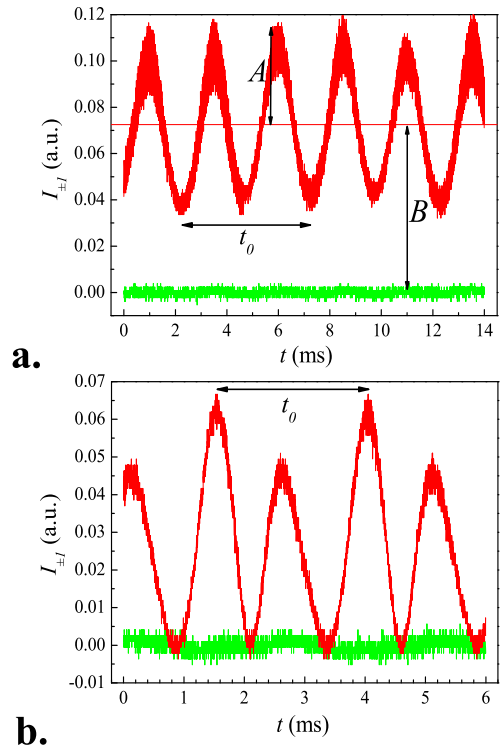}
\caption{(Color online) Same as in Figure~\ref{fig:stacsEC},
measured in samples exhibiting traveling s-EC patterns. (a) Phase 5,
$d=3.4\mu$m, $T=30^{\circ}$C, $f=200$Hz, traveling conductive s-EC,
$A/B \approx 0.57$; (b) Phase 4, $d=3.2\mu$m, $T=30^{\circ}$C,
$f=400$Hz, traveling dielectric s-EC, $A/B=1$.}\label{fig:travsEC}
\end{figure}

In Phase 5, below $f_c$, stationary as well as traveling conductive
s-EC rolls could be seen which gave an opportunity for an
experimental testing of the case. The frequency range in which the
traveling conductive s-EC rolls occur, the Hopf frequency, as well
as the spatial homogeneity of the pattern increases as the thickness
of the sample is decreased. Fig.~\ref{fig:travsEC}a shows an example
of diffraction measurements on a thin, $d = 3.4\mu$m sample of Phase
5, where spatially homogeneous traveling (with high speed)
conductive s-EC rolls have been observed. As one sees, the $2f$
intensity modulations here are considerably larger ($A/B \approx
0.57$) than in the case of stationary conductive s-EC
(Fig.~\ref{fig:stacsEC}a).

We note here, that for the traveling conductive s-EC the
diffraction fringes became more diffuse, indicating less spatial
regularity of the pattern compared to the stationary one. This is,
however, not expected to affect the $A/B$ ratio.

While traveling conductive s-EC rolls are commonly observed,
traveling dielectric s-EC rolls have not been reported until
recently \cite{Toth08}. If one follows the predictions of the WEM
and the experimental findings regarding the conditions under which
traveling conductive s-EC rolls are observed, one expects that for
the occurrence of traveling dielectric s-EC patterns a compound with
low electrical conductivity is needed in a very thin sample.
Therefore, we have prepared a $d=3.2\mu$m thick sample of Phase 4,
in which traveling dielectric s-EC rolls could indeed be detected.
For this pattern $I_{\pm 1}(t)$ exhibited a $2f$ modulation of the
intensity as illustrated in Fig.~\ref{fig:travsEC}b. The intensity
decreases to the background level in each half period, i.e. the
curve yields $A/B=1$, similarly to that measured for stationary
dielectric s-EC (Fig.~\ref{fig:stacsEC}b).

The measurements above have proven that the Hopf-bifurcation has no
effect on the $A/B$ ratio in the dielectric regime of s-EC; however,
in the case of traveling conductive s-EC rolls the intensity
modulation (as shown in Fig.~\ref{fig:travsEC}a) becomes much larger
than what would be predicted by the SM and has been measured for
stationary conductive s-EC (Fig.~\ref{fig:stacsEC}a).

\begin{figure}[h!t]
\includegraphics[width=8cm]{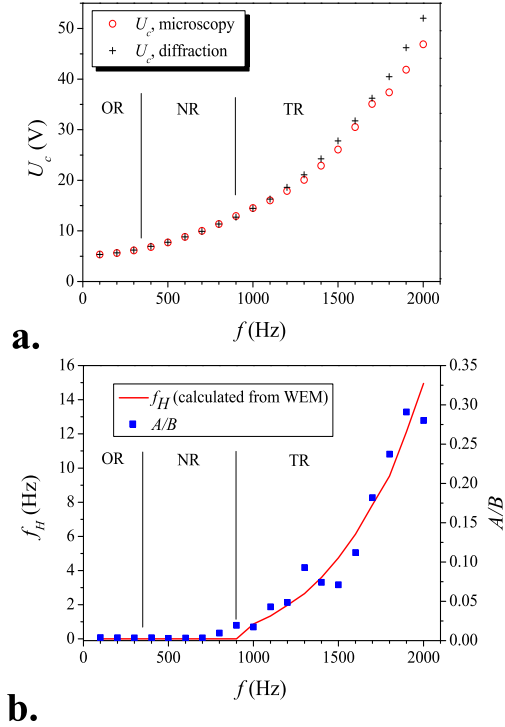}\\
\caption{(Color online) {\bf (a.)} Frequency dependence of the onset
voltage $U_c$ of conductive s-EC rolls determined both with optical
microscopy (open circles) and with diffraction (pluses); {\bf (b.)}
the ratio $A/B$ (solid squares) of the $2f$-modulated ($A$) and the
constant ($B$) part of the diffracted fringe intensity measured in a
mixture of Phase 5 and Phase 5A (7:1) at $T=30^\circ$C, as well as
the frequency dependence of the Hopf-frequency $f_H$ calculated from
the WEM (solid line). The frequency ranges of the stationary oblique
(OR), stationary normal (NR) and traveling normal (TR) rolls are
separated by vertical lines.} \label{fig:AoverBvf}
\end{figure}

This astonishing behaviour justifies a more detailed study of the
problem in an NLC where both stationary and traveling conductive
s-EC rolls can be detected in a conveniently broad frequency range.
For this purpose a mixture of Phase 5 and Phase 5A has been prepared
in a mass ratio of 7:1. In Fig.~\ref{fig:AoverBvf}a the $U_c(f)$
curves are plotted for a $d=10.3 \mu$m thick sample (determined
independently both with polarizing microscopy and by diffraction
measurements). It is seen, that the composition of the mixture
assured broad frequency ranges for stationary oblique (OR) and
normal (NR) rolls as well as for traveling (TR) roll patterns. The
frequency dependence of the $A/B$ ratio of $I_{\pm 1}(t)$ is shown
in Fig.~\ref{fig:AoverBvf}b as symbols. Obviously, $A/B$ remains
very small in the whole frequency range of stationary patterns
(independent of whether OR or NR), however, it starts to grow with
$f$ considerably when entering the range of traveling roll scenarios
(with the Hopf frequency $f_H$ also increasing with $f$). Using the
WEM one can calculate the frequency dependence of $f_H$
\cite{Treiber95}. Results of these calculations are also presented
in Fig.~\ref{fig:AoverBvf}b by the solid line. We want to emphasize
that this is {\it not a fit}; besides the known material parameters
and the sample thickness, only the measured threshold voltages
$U_c(f)$ and the frequency at which the rolls start to travel
($f=900$Hz) are required for the calculations \cite{Treiber95}.
Fig.~\ref{fig:AoverBvf}b clearly demonstrates that the ratio $A/B$
scales with $f_H$.

Finally, we mention that in the high temperature range ($0.85
\lessapprox T^{*} < 1$) of the nematic phase compound {\bf 8/7}
possesses the $(-,+)$ parameter combination  and consequently, s-EC
occurs \cite{Toth07}. In this high temperature range, traveling
conductive s-EC rolls have been detected below $f_c$, while
stationary dielectric s-EC rolls above it by polarizing microscopy.
Our light diffraction experiments yielded $A/B \sim 0.1$ for the
patterns below $f_c$, while above $f_c$ we obtained $A/B=1$; i.e.,
the intensity modulations found in the s-EC of {\bf 8/7} are in full
agreement with those obtained for other nematics having traveling
conductive s-EC (see, e.g., Figs.~\ref{fig:travsEC}a and
~\ref{fig:AoverBvf}b) and stationary dielectric s-EC
(Fig.~\ref{fig:stacsEC}b).

\subsection{\label{subsec:8_7nsEC}Light diffraction on ns-EC patterns}

The Carr-Helfrich mechanism, on which the SM is based, is not
capable to explain the ns-EC patterns occurring in $(-,-)$ NLCs. It
has been shown, however, recently that a proper theoretical
description can be provided by incorporating flexoelectricity as the
source of destabilization into the SM \cite{Krekhov08}. This
extended SM concludes that flexoelectricity establishes a coupling
between the conductive and dielectric modes, and that the director
should oscillate with the frequency of the driving voltage similarly
to the case of dielectric s-EC, however, $\mathbf{q_c}$ is different
and the spatial director distribution is more complex. The director
has a spatially alternating tilt and azimuthal modulation which
oscillate in phase \cite{Krekhovunpub}; therefore at one instant in
each half period there is no deformation. As a consequence, one
would expect that for this pattern morphology the diffracted
intensity should fall down to the background level twice in each
period, just as in Figs.~\ref{fig:stacsEC}b or \ref{fig:travsEC}b.

\begin{figure}[h!t]
\includegraphics[width=8cm]{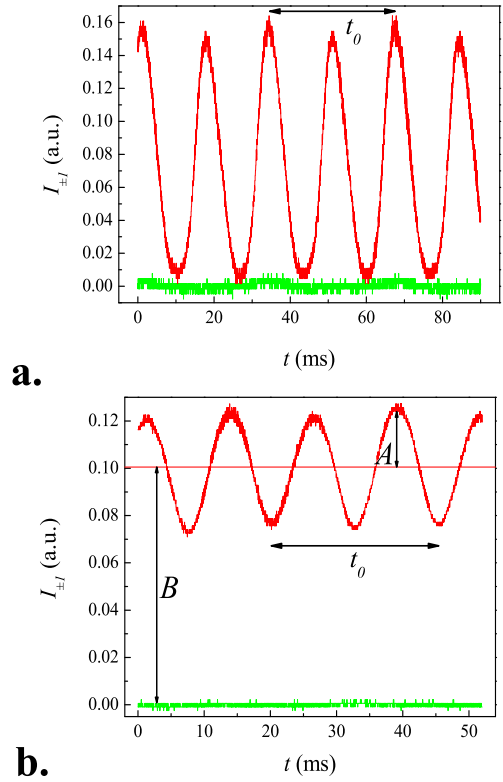}
\caption{(Color online) Same as in Figure~\ref{fig:stacsEC},
measured in samples of (a) {\bf 10/6}, where stationary ns-EC patterns
appear at the onset of convection, $d=11\mu$m, $T^{*} \approx 0.6$,
$f=30$Hz, $A/B=1$; (b) {\bf 8/7} in ($d$, $T^{*}$, $f$) parameter
range where traveling ns-EC patterns appear at the onset of convection,
$d=13.2\mu$m $T^{*}=0.38$ $f=40$Hz, $A/B \approx 0.25$.}\label{fig:8_7nsEC}
\end{figure}

Previous observations with polarizing microscopy on the compound
{\bf 10/6} \cite{Ela} have shown that in a broad temperature range
of the nematic phase stationary ns-EC appears at the onset of the
instability. Traveling ns-EC pattern has not been observed on this
compound. Fig.~\ref{fig:8_7nsEC}a shows the temporal dependence of
the intensity $I_{\pm 1}(t)$ of the first order diffraction fringe
just above the onset of the stationary ns-EC instability measured
in a $d=11\mu$m thick sample of {\bf 10/6} at $T^{*} \approx 0.6$
and $f=30$Hz. Clearly $A/B=1$ fulfils, thus confirming in this
respect the "dielectric time symmetry" of the stationary ns-EC
pattern.

Nonstandard EC rolls may also be traveling, as it was observed in
the low temperature range ($T^{*} \lessapprox 0.85$) of the
nematic phase of {\bf 8/7} where the substance is of $(-,-)$ type
\cite{Toth07}. Fig.~\ref{fig:8_7nsEC}b represents a diffraction
measurement in this low temperature range taken at $T^*=0.38$ and
$f=40$Hz using a $d=13.2\mu$m thick sample. It follows from the
figure that $A/B \approx 0.25$. It was found in general, that by
varying $d$, $T^*$ and/or $f$ the $A/B$ ratio falls into the range
of $0.1 - 0.5$ for traveling ns-EC (i.e., into the same range as
measured for the traveling conductive s-EC discussed in
Sec.~\ref{subsec:travsEC}). This indicates that for traveling
ns-EC patterns the "dielectric time symmetry" seems to fail.

\subsection{\label{subsec:5CB}Light diffraction on the "roll pattern" of a $(+,+)$ nematic}

Very recently two EC pattern morphologies -- a "cellular" and a
subsequent "roll" pattern -- have been detected in the same
frequency range in a $(+,+)$ nematic, namely in 5CB
\cite{Kumar2010}. As the mechanisms of formation of these patterns
still wait for exploration, no theoretical guesses can be made for
their spatio-temporal director distribution.

Unfortunately, light diffraction experiments on the cellular
pattern did not provide any information about the temporal
evolution of the light intensity $I_{\pm 1}$. The most probable
reasons for this are the low contrast and the relatively large
wavelength of the pattern (see Figs. 2b and 4b of Ref.
\cite{Kumar2010}), which makes the first order diffraction fringe
positioned too close to the central beam where the intensity of
the scattered light (the background) is too high. Therefore, using
a high-speed camera could be a more powerful tool for studying the
temporal evolution of the cellular pattern. These studies are
currently in progress.

\begin{figure}[h!t]
\includegraphics[width=8cm]{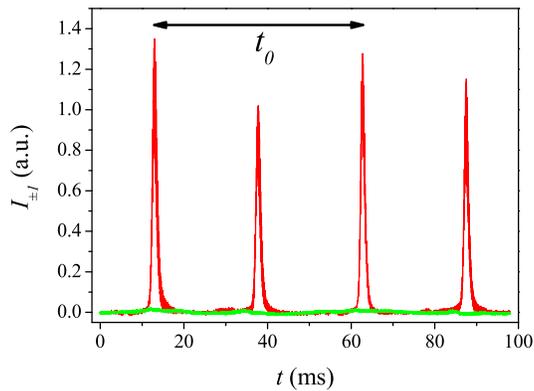}
\caption{(Color online) Same as in Figure~\ref{fig:stacsEC},
measured in a $(+,+)$ sample of 5CB below (at $U=20$V) and above (at
$U=50$V) the onset of the roll pattern \cite{Kumar2010}.
$d=19.5\mu$m, $T=30^{\circ}$C, $f=20$Hz.}\label{fig:5CB}
\end{figure}

For the roll pattern (the secondary instability which has a much
lower wavelength than the cellular one) near its onset ($U_c
\approx 38$V in a $d=19.5\mu$m thick planar cell at $f=20$Hz and
$T=30^{\circ}$C) $I_{\pm 1} (t)$ has been reported as a sequence
of bursts appearing twice within one period of the driving voltage
\cite{Kumar2010}. Interestingly, the roll pattern remains stable
over a wide range of the driving voltage. In Fig.~\ref{fig:5CB}
the temporal evolution $I_{\pm 1} (t)$ is depicted much below (at
$U=20$V) and much above $U_c$ (at $U=50$V). As one sees, even
highly above the threshold (in terms of the dimensionless control
parameter $\varepsilon = (U/U_c)^2-1 >0.7$), the temporal
evolution of the pattern remains the same as that at the onset
({\it cf.} Fig. 5 in Ref. \cite{Kumar2010}): two "short-time"
bursts within one period of the driving voltage, and no pattern in
most of the time.

\section{\label{sec:discussion}Discussion and summary}

Our experimental results presented above clearly show that
monitoring the oscillations of the light intensity $I_{\pm 1}(t)$
diffracted from various EC patterns is a convenient tool for the
analysis of temporal behaviour and is a selective method for
distinguishing certain pattern morphologies. We have managed to
prove that for those patterns where the proper theoretical
description is well developed (stationary s-EC and ns-EC), the
experimental results are in full agreement with the theoretical
predictions. It has, however, turned out that the temporal behavior
of traveling patterns is mostly significantly different from those
of stationary ones and there is a correlation between the magnitude
of the modulation and the Hopf frequency characterizing the
traveling.

In view of the fact that the frequency of the illuminating light
($\approx 10^{14}$Hz) is several orders of magnitude larger than the
Hopf frequency ($f_H < 20$Hz) or even the driving frequency ($f <
2000$Hz), there is no reason to assume that diffraction optics could
be responsible for the altered time dependence of the traveling
rolls. Rather one can state that the diffraction fringe pattern
appearing on the screen corresponds to a snapshot in the
Fourier-space of the momentary director distribution.

Therefore we conclude that the spatio-temporal director distribution
in traveling roll patterns should be significantly different
compared to that of the corresponding stationary patterns. The
resulting difference in the intensity modulation may occur via at
least two ways. First, trivially, an increase of the weight of the
higher ($2f$) harmonic in $\theta(t)$ should induce a larger
modulation (larger $A/B$). Second, it has been shown recently
\cite{Pesch06} that the diffraction efficiency of a pattern depends
also on the $z$-profile of the director. Thus if $\theta(z)$ changes
significantly moving from stationary to traveling rolls, it could
also be responsible for a change of $A/B$, even if the harmonic
weights remain the same. A thorough theoretical/numerical analysis
of the problem by the WEM would presumably clarify the situation,
and give some clue to the correlation between $A/B$ and the
Hopf-frequency which exists according to Fig.~\ref{fig:AoverBvf}.

In case of traveling dielectric rolls no significant change has been
detected compared to the case of stationary dielectric rolls. The
intensity diminishes to zero twice in each period, confirming that
the director crosses the $\theta =0$ position. A possible change in
the weights of higher harmonics would not change this behaviour;
that could show up only in higher harmonic modulation of the
diffracted intensity which has not been investigated here.

We note that while only $2f$ modulations have been expected in
$I_{\pm 1}$, actually a non-negligible basic harmonic modulation
could also be detected, as it is noticeable in
Figs.~\ref{fig:stacsEC}b and \ref{fig:travsEC}b. The $f$ modulation
actually means a polarity sensitivity of the response. We propose
that this feature might be related to flexoelectricity. While
according to the SM (flexoelectricity is neglected) the different
time symmetries of the conductive and the dielectric regimes are
clearly reflected by the parity of the time harmonics, it has been
shown recently \cite{Krekhov08} that in the extended SM
(flexoelectricity included) there is a coupling between modes;
therefore the harmonic expansion may have even and odd harmonics as
well. If the director has both a constant and a basic harmonic
component, the $I_{\pm1} \propto \theta^2$ intensity should have a
basic harmonic component too.

Diffraction on ns-EC patterns is a more delicate problem. While for
stationary ns-EC the experimental findings seem to agree with the
theoretical predictions, in case of traveling ns-EC the temporal
evolution of the diffraction intensity alters qualitatively. The
intensity does not fall down to the background level; $A/B$ reduces
from 1 to about $0.1-0.5$ which is approximately the same value as
that for traveling conductive s-EC rolls. It has also been found by
optical microscopy that whenever the $A/B$ ratios of the two kinds
of patterns are similar, the Hopf-frequencies are also of the same
order of magnitude. Understanding that the diffraction intensity is
related to a snapshot of the director field, one might conclude that
in traveling ns-EC there is no such time instant when the director
returns to its initial undeformed state. That might, e.g., occur if
the tilt and azimuthal oscillations (which are in phase for
stationary ns-EC) become phase shifted due to the traveling.
Theoretical description of this phenomenon is a great challenge; it
would require an extension of the WEM by inclusion of
flexoelectricity.

Finally, light diffraction experiments on the $(+,+)$ nematic (5CB)
have revealed a completely different temporal evolution of the light
intensity. Instead of $2f$ intensity oscillations (which were
observed both in s-EC and in ns-EC), here we have detected a
sequence of bursts occurring twice within one period, while in most
of the time the pattern does not exists at all. Bursts may exist in
s-EC too: it has been reported for dielectric s-EC \cite{May08} and
has recently been detected in the conductive regime as well
\cite{Eberunpub}. Surprisingly, however, the bursts in 5CB occur
even at quite high frequencies (around 20Hz, see
Fig.~\ref{fig:5CB}), while in s-EC (Phase 4 or Phase 5) they become
observable only at very low $f$ ($<1$Hz) driving. This huge
difference in frequency range cannot be explained by differences in
the sample thickness or of the viscosity coefficients. Also, in the
$(+,+)$ NLC the phase shift between the bursts and the driving
voltage appears to be independent of $f$, which is not the case in
the dielectric s-EC \cite{May08}. These observations suggest that
the instability mechanism in $(+,+)$ nematics is completely
different from that of the roll patterns of either s-EC of $(-,+)$
or ns-EC of $(-,-)$ NLCs, and the phenomenon is governed -- at least
partially -- by different time scales.

\section*{ACKNOWLEDGEMENTS}

We thank to W. Pesch and A.P. Krekhov for fruitful discussions.
Financial support by the Hungarian Research Fund Contract No. OTKA-K81250
is gratefully acknowledged.

\end{document}